# Deep Learning-based Assessment of Hepatic Steatosis on chest CT

Zhongyi Zhang[1,2] ; Jakob Weiss[1,2,3]; Jana Taron[3]; Roman Zeleznik; Michael T. Lu; Hugo J.W.L. Aerts *

1. Artificial Intelligence in Medicine (AIM) Program, Mass General Brigham, Harvard Medical School, Boston, MA, USA
2. Department of Radiation Oncology, Dana-Farber Cancer Institute and Brigham and Women's Hospital, Harvard Medical School, Boston, MA, USA
3 Department of Diagnostic and Interventional Radiology, University Medical Center Freiburg, Freiburg, Germany

* Corresponding author
Hugo Aerts, PhD
Director, Artificial Intelligence in Medicine (AIM) Program, Mass General Brigham, Harvard Medical School, Boston, MA
P - 617.525.7156, F - 617.582.6037, Email: haerts@bwh.harvard.edu

**Key Points**
1. We developed a 3D deep learning model to automatically and robustly segment the liver on non-contrast enhanced chest CT scans.
2. The deep learning-based liver fat quantification was highly correlated with the measurements of expert readers on the test set of 980 CT scans.
3. The end-to-end automatic assessment for hepatic steatosis showed a high agreement with expert readers at high speed and low costs.

## ABSTRACT

**Purpose:** Automatic methods are required for the early detection of hepatic steatosis to avoid progression to cirrhosis and cancer. Here, we developed a fully automated deep learning pipeline to quantify hepatic steatosis on non-contrast enhanced chest computed tomography (CT) scans.

**Materials and Methods:** We developed and evaluated our pipeline on chest CT images of 1,431 randomly selected National Lung Screening Trial (NLST) participants. A dataset of 451 CT scans with volumetric liver segmentations of expert readers was used for training a deep learning model. For testing, in an independent dataset of 980 CT scans hepatic attenuation was manually measured by an expert reader on three cross-sectional images at different hepatic levels by selecting three circular regions of interest. Additionally, 100 randomly selected cases of the test set were volumetrically segmented by expert readers. Hepatic steatosis on the test set was defined as mean hepatic attenuation of < 40 Hounsfield unit. Spearman correlation was conducted to analyze liver fat quantification accuracy and the Cohen's Kappa coefficient was calculated for hepatic steatosis prediction reliability.

**Results:** Our pipeline demonstrated strong performance and achieved a mean dice score of 0.970 ± 0.014 for the volumetric liver segmentation. The spearman correlation of the liver fat quantification was 0.954 (P <0.0001) between the automated and expert reader measurements. The cohen's kappa coefficient was 0.875 for automatic assessment of hepatic steatosis.

**Conclusion:** We developed a fully automatic deep learning-based pipeline for the assessment of hepatic steatosis in chest CT images. With the fast and cheap screening of hepatic steatosis, our pipeline has the potential to help initiate preventive measures to avoid progression to cirrhosis and cancer.

**Abbreviations:** 3D = Three-dimensional, CI = Confidence interval, HS = Hepatic steatosis, HU = Hounsfield units, ICC = Intraclass coefficient, NLST = National Lung Screening Trial, ROI = Region of interest, std = Standard deviation.

**Summary:** Fully automatic volumetric assessment of hepatic steatosis is feasible in a lung cancer screening eligible population and might help to initiate preventive measures to avoid progression to cirrhosis and cancer.

**INTRODUCTION**

Hepatic steatosis (HS) is defined as the intrahepatic fat accumulation of at least 5% (1). It is a public health concern with an estimated prevalence of 20% to 30% in the western population (2). Nonalcoholic fatty liver disease (NAFLD), a potentially reversible condition, is the most common manifestation of HS and poses the risk for progression to liver cirrhosis and cancer (3). Thus, fast, cost-effective, and safe methods are desirable to improve early diagnosis of HS to reduce morbidity and mortality. Liver biopsy is the current reference standard for HS diagnosis (4) which is associated with high costs, sampling error, and potentially life-threatening complications making population-based screening for HS impossible (4,5).

Non-contrast enhanced CT has been proposed as a non-invasive alternative for the diagnosis of HS (6). In particular, a threshold of 40 Hounsfield units (HU) in regions of interest (ROI) of normally appearing liver parenchyma with sparing of large vessels and bile ducts has been established for the diagnosis of HS on non-contrast enhanced CT scans (7). However, the ROI-based approach for HS detection is time-consuming and requires special experience and covers only a relatively small area of the entire liver, thus, regional HS may go unnoticed similar to biopsy-related sampling error. Therefore, alternative methods are needed to save time and reduce potential misdiagnosis due to sampling problems.

Recent advances in artificial intelligence might provide a solution to this problem. Deep learning, a sub-branch of artificial intelligence, has shown high performance for identifying and segmenting objects within the three-dimensional image space including liver segmentation (8,9). Usually, liver segmentation is performed on abdominal imaging studies as they fully cover the liver (10). Deep learning-based HS quantification from non-contrast enhanced chest CT scans has not been reported before although major parts of the liver are usually within the field of view.

Here, we developed and evaluated a deep learning algorithm to automatically quantify HS on non-contrast enhanced chest lung cancer screening CT's. The accuracy was compared to the currently used manual ROI-based method conducted by expert readers Fully automatic assessments of hepatic steatosis might help to initiate preventive measures to avoid progression to cirrhosis and cancer.

**METHODS**

*Development of the Deep learning-based pipeline.* In this study, a deep-learning pipeline (Fig. 1A) was developed and tested from June 2020 to Dec 2021 for volumetric liver segmentation and hepatic steatosis assessment. The pipeline consisted of two steps: 1) volumetric liver segmentation by deep learning model, 2) hepatic steatosis assessment.

**Figure 1. A:** The deep-learning pipeline was developed and tested on 1,451 computed tomography (CT) scans from NLST. ROI-based attenuation was quantified for all of 980 test scans, among which 100 random scans were volumetrically segmented to assess the model segmentation. **B:** The accuracy of the framework was evaluated with respect to liver segmentation and liver fat quantification. **C:** Liver fat quantification was measured in Hounsfield units in three independent ways: 1. The mean

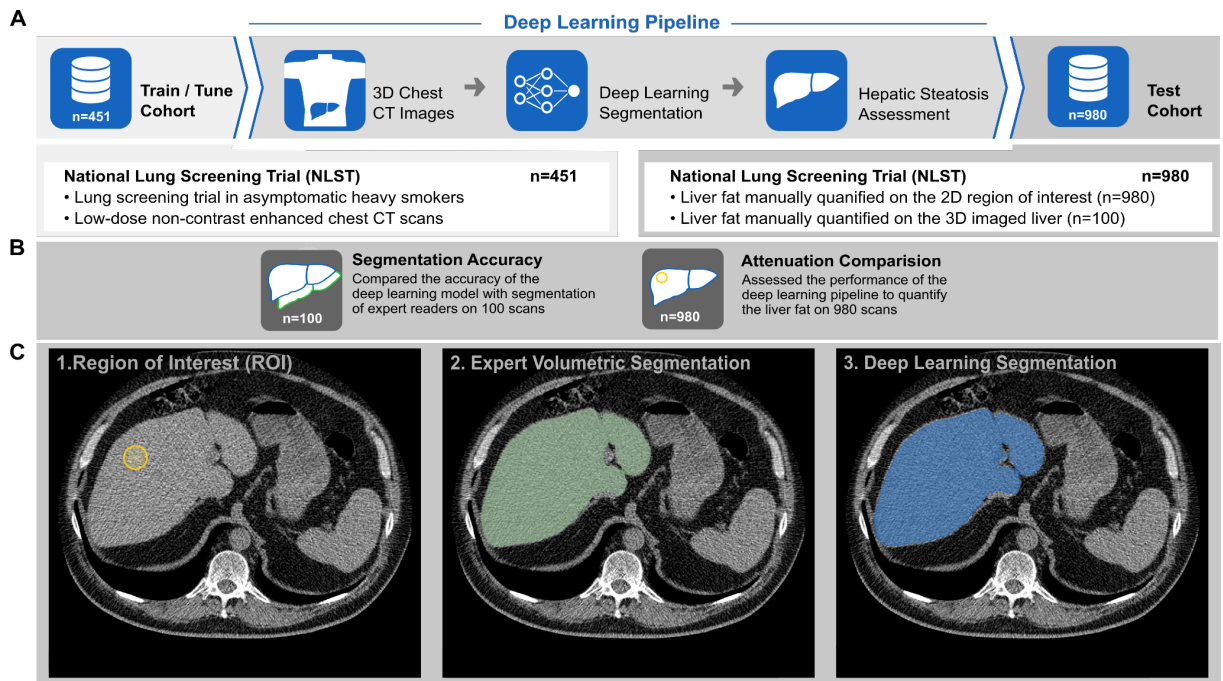

attenuation of the region of interest (ROI) on three cross-sectional images at different hepatic levels; 2. The mean attenuation of manual volumetric segmentation by expert readers; 3. The mean attenuation of deep learning-based volumetric segmentation.

*Study population.* Study materials were selected from non-contrast enhanced chest CT scans of the National Lung Screening Trial (NLST), a randomized controlled trial for lung cancer screening (11). 53,454 asymptomatic heavy smokers were enrolled at 33 medical centers across the United States of America in the NLST. A total of 26,722 participants were randomly assigned to the CT group undergoing non-contrast enhanced low-dose chest CT between 2002 and 2007 using a variety of CT scanners. Trial consent was provided by the institutional review board at each site.

Our study was approved to include 15,000 randomly selected participants from the full NLST cohort. CT scans were chosen per participant at the baseline time point (T0) with soft kernel preferred over hard kernel reconstructed images.

**Figure 2:** The consort diagram for the data partition of our study cohort (n=15,000). A random subset of 1,431 scans was chosen for the development and independent testing of the deep learning model. For testing, the HS of participants was

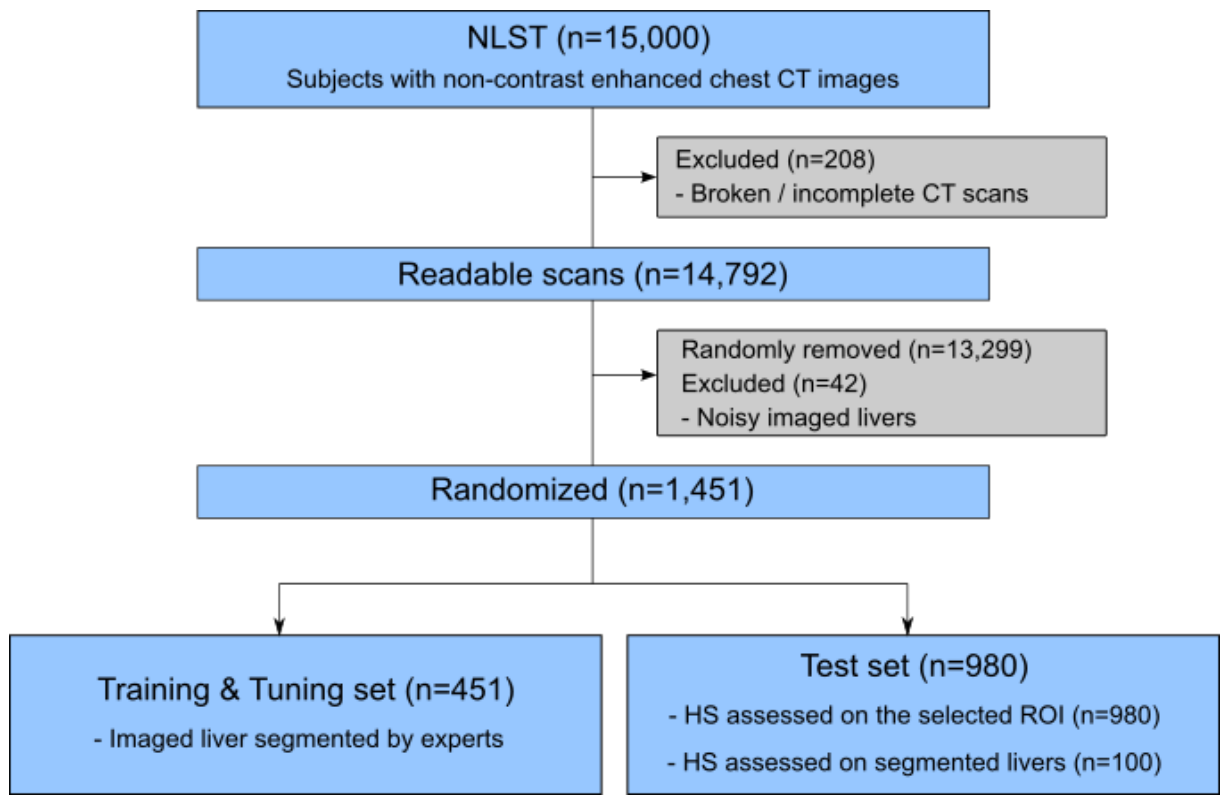

quantified by expert readers in an ROI-based way. In addition, a random subset (n=100) was also volumetrically segmented to assess the model performance. **NLST**: National Lung Screening Trial; **HS**: Hepatic steatosis; **ROI**: region of interest

We excluded 268 participants, where 208 subjects had a broken or incomplete scan, and 60 subjects had a noise-imaged liver that failed to meet our quality requirements. A random subset of 1,431 scans was chosen as the final developing and testing cohort as shown in Fig. 2. Among those, HS was manually quantified via ROI-based measurements in a random sample of 980 participants (Fig. 1C). In 551 additional participants, the liver was volumetrically segmented by expert readers. Among them, 451 randomly selected CT scans were taken as the training and tuning cohort and 100 independent scans were treated as the test data.

*HS Annotation by Expert Readers.* HS was manually quantified by measuring the mean liver attenuation using two different methods (Fig. 1C).

First, hepatic attenuation was measured on three cross-sectional images at different hepatic levels by selecting three circular ROIs with a minimal area of 2 cm$^2$. All measurements were performed by a board-certified radiologist (J.T.) with 7 years of experience using an open-source software package (3D Slicer version 4). Non-parenchymal structures like hepatic veins or bile ducts, focal nodules, or other types of heterogeneity were spared (12). The hepatic attenuation was calculated as the mean of those three ROIs.

For the second method, the liver was volumetrically segmented using 3D Slicer version 4 as well. The manual volumetric segmentations were used for the deep learning model development. The segmentations were performed by a board-certified radiologist (J.W.) with 6 years of experience.

For both methods, HS was defined as a mean hepatic attenuation of < 40 HU as previously described (7,13).

*HS quantification by the Deep Learning-based pipeline.* We developed a deep learning model to volumetrically segment the liver in non-contrast enhanced chest CT scans. All voxels within the segmented liver were analyzed to compute the mean volumetric attenuation. The deep learning model used a U-Net architecture which was developed for medical image segmentation tasks and has shown good results in several published studies (14,15). Our development cohort consisted of 451 CT scans, randomly selected from NLST and split into the training (n=345) and tuning (n=106) cohort. The testing set consisted of 100 held-out CT scans not seen during any part of the training.

The liver was first roughly segmented in chest CTs using a method (10) that was designed for liver segmentation of abdominal CTs. Rough initial segmentations were then corrected in all subjects. The three-dimensional input scans for our model were padded or cropped to 512 x 512 pixel (px) in the X and Y dimension and resampled to an uniform resolution of 0.7 x 0.7 x 2.5 mm/px. Further details of the model development are provided in the supplemental methods.

At last, the largest connected component in the model segmentation was retained by the connected-component analysis. Training, tuning, and testing were done on a Linux workstation using Pytorch version 1.6.0 (16) and MONAI framework version 0.4.0 (17).

*Statistical Analysis.* The accuracy of deep learning segmentation was evaluated using the Dice coefficient, average symmetric surface distance (ASSD), Hausdorff distance, and Jaccard distance (Python package: medpy.metric.binary, V0.4.0). To assess the correlation between manual and automatic measurements, Spearman's correlation coefficient (18) (Python package: scipy.stats.spearmanr, version 1.5.3) and the intraclass coefficient (ICC) (19) (SPSS statistics software version 26, one-way random absolute agreement Single Measures) were calculated. The pipeline accuracy for HS assessment was rated using Cohen's Kappa coefficient (20) (Python package: sklearn.metrics.cohen_kappa_score, version 0.23.2) and the concordance rate (calculated as the number of concordant pairs divided by the number of all pairs).

**RESULTS**

We developed a fully automatic deep learning-based pipeline to volumetrically segment the liver and quantify hepatic steatosis based on mean liver attenuation. Figure 1 summarized the pipeline's workflow. Baseline demographic and clinical risk factors for the study participants were as follows were: age is 61.5 ± 5.1 (mean ± std); women 40.9%, 9.5% of individuals have diabetes; 35.6% have arterial hypertension; body mass index is 27.9 ± 5.1 (kg/m$^2$).

*Comparison of deep learning-based Automatic and Manual ROI-based measurements.* To assess the deep learning system's performance for HS quantification, we compared the deep learning quantified automatic volumetric attenuation to manual ROI-based liver attenuation in our test cohort (n=980).

**Figure 3. A**: Scatter plot shows mean liver attenuation measured in HU by the deep-learning algorithm and expert readers.

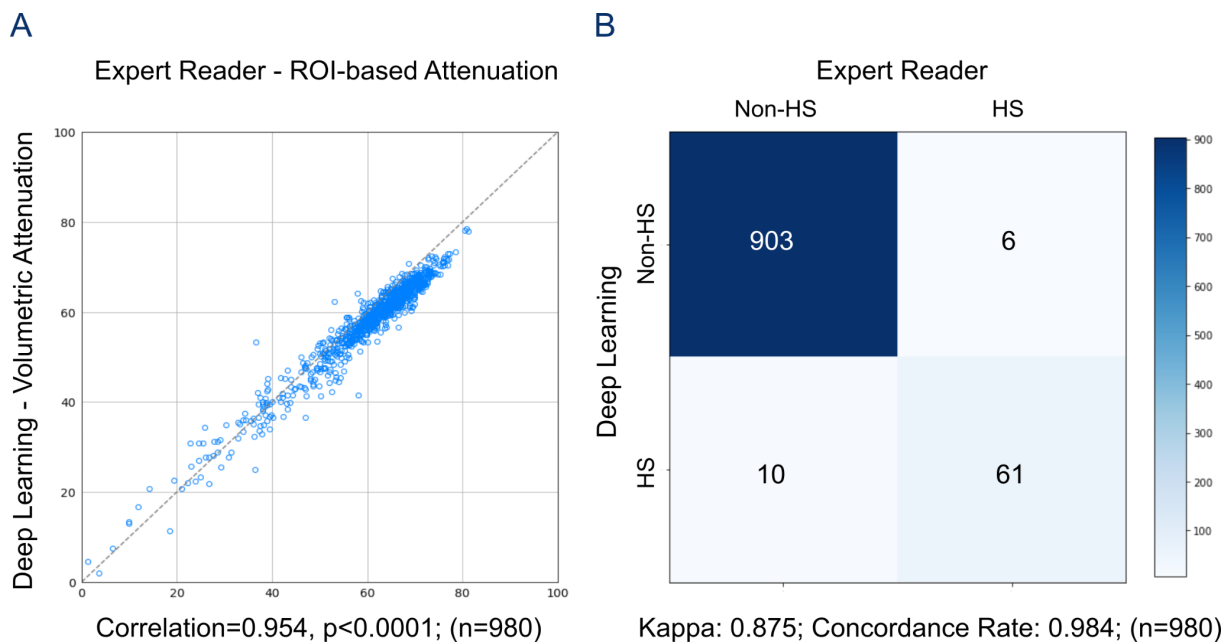

Deep learning-based attenuation is based on automatic volumetric segmentation of the imaged liver. Expert readers quantified the liver attenuation based on the ROI of liver slices. **B**: A confusion matrix was conducted to compare the HS classification quantified by deep learning and expert readers. Non-HS: >= 40 HU; HS: <40 HU. **HS**: hepatic steatosis; **HU**: Hounsfield units; **ROI**: region of interest.

There was a high spearman's correlation between auto and manual of 0.954 (P<0.0001) as shown in Fig. 3A. The ICC is 0.943 with a 95% CI from 0.936 to 0.950 between two groups of attenuation measurements. Deep learning-based HS classification achieved a Cohen's kappa coefficient of 0.875 and a concordance rate of 0.98 as shown in Fig. 3B. The accuracy of deep learning classification in the confusion matrix was 0.984; precision was 0.910; recall was 0.860; F1 score was 0.884.

*Comparison of Automatic deep learning and Manual Volumetric measurements.* To evaluate the performance of the deep learning system for volumetric liver fat quantification, we compared the automatic and expert reader measurements in the independent test set (n=100).

**Figure 4:** Volumetric liver mean attenuation measured in HU and liver volume measured in liter: Artificial Intelligence vs Expert

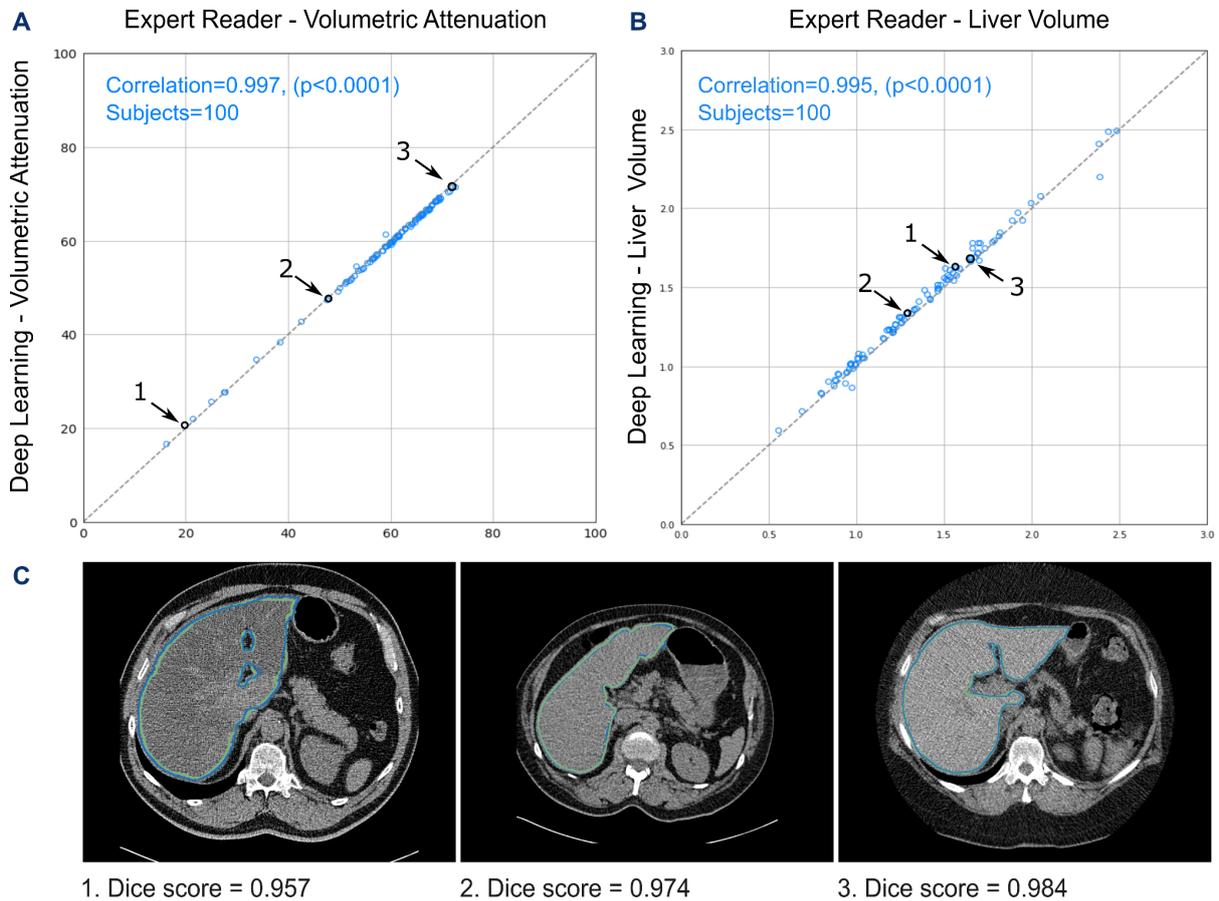

1. Dice score = 0.957    2. Dice score = 0.974    3. Dice score = 0.984

Reader, along with three examples of segmentation comparisons. In Figures A and B, Spearman correlation coefficients with associated two-sided p-value were estimated for the comparison of expert readers and artificial intelligence measurements in HU and liter. For Figure C.1, 2, 3, each example image was sliced from the CT scan with the AI (blue) and expert reader (green) segmentation masks.

The deep learning-based segmentation achieved a high overlap with manual segmentations in 100 random NLST scans: 0.970 ± 0.014 for mean Dice similarity coefficient (DSC); 0.802 ± 0.494 for average symmetric surface distance (ASSD); 17.2 ± 11.3 mm for the Hausdorff distance; 0.943 ± 0.025 (mean ± std) for the Jaccard distance. Representative segmentation examples are shown in Figure 4. Incorrectly segmented cases are shown in Supplemental Figure 2. All voxels within the segmented liver were analyzed to compute the mean volumetric liver attenuation and volume. The automatic attenuation and volume had a high correlation of 0.997 and 0.995 (p<0.0001 for both) with the manual expert reader assessments (Fig. 4A, B). The ICC is 0.999 between deep learning and manual volumetric attenuation along with the 95% CI from 0.998 to 0.999. ICC between different quantifications of volume is 0.992, along with the 95% confidence interval from 0.988 to 0.994.

**DISCUSSION**

In this study, we demonstrate the feasibility of a deep learning pipeline for opportunistic screening of HS in non-contrast enhanced CTs of the National lung screening trial. Our deep learning model achieved high accuracy for the 3D liver segmentation. Built upon it, our pipeline reached an excellent correlation with expert readers for liver fat quantification and HS detection. With the huge number of participants recommended for lung cancer screening (21), our pipeline has the potential to contribute to HS detection in lung cancer screening eligible individuals.

Traditionally, HS is quantified on abdominal non-contrast enhanced CTs by manual ROI placements (22), which requires time and professional experience. Also, ROI-based HS quantification may bring sampling errors since it covers only a relatively small area of the liver. In contrast, our pipeline has several innovations: first, we demonstrated the feasibility of volumetric assessment. Our pipeline might handle the sampling error since the deep learning-based liver attenuation was calculated volumetrically. Second, Our pipeline can assess a new CT scan in under two seconds without human input, which makes it an 'end-to-end' system at high speed and low additional costs. Third, the validated deep learning pipeline is shared with the public for accelerated adoption by both academic and commercial entities (23).

To the best of our knowledge, this study is the first work that uses deep learning for 3D liver segmentation and volumetric HS quantification on non-contrast enhanced chest CTs. Among imaging methods, MRI is the current reference standard for non-invasive HS diagnosis thanks to its high accuracy (24,25). However, MRI is expensive and not broadly available (26). With the big number of participants for the lung cancer screening (21), it was a logical step to assess the potential of the prognostic value of non-contrast enhanced CT scans. A recent study proposed a fully automatic deep learning tool for liver segmentation and attenuation assessment on the non-enhanced abdominal CT (10). However, this tool is not feasible for lung screening chest CTs as it was designed for the quantification of abdominal CTs only. Another study proposed an automatic method for liver density quantification on the non-enhanced chest CTs (27). However, this method was not highly generalizable as it assumed the liver was fairly homogenous and did not exceed the region below the right lung. Our pipeline made no assumptions to the imaged liver. Furthermore, several deep learning models for liver segmentation in contrast-enhanced only CTs have been published (9). The best model of the challenge achieved a high accuracy of the dice score around 0.97. Since the liver at CT scans was imaged in different contrasts from our study materials, their performances were not comparable to ours.

There are several reasons for the high performance of the proposed pipeline. The observed high correlation could be attributed to the advantage of the model architecture. A review article has shown the dominant performance of the U-Net architecture in many medical image segmentation tasks (28). Another possible explanation for the observed result is the vast amount of training data from various clinical sites. Research has shown that the model would be more robust and generalizable with more training samples (29).

We also acknowledge certain limitations of our study. The model was developed and validated on chest CT, in which imaged livers are incomplete in most scans. Therefore, the mean attenuation quantified of the imaged liver might be slightly different from the mean density of the whole liver. Also, more external datasets would be beneficial to prove the generalizability of our pipeline. Moreover, non-enhanced CT has some shortcomings as an imaging modality to detect HS (30,31). For example, liver attenuation measurements could vary between CT scanners and reconstruction algorithms (32) and be influenced by some factors other than fat, such as the presence of metallic ions (33).

In conclusion, we proposed a fully automatic deep learning pipeline for volumetric assessment of HS on chest CTs in the lung cancer screening eligible population. With the fast and cheap screening of HS, our end-to-end pipeline could help to initiate preventive measures to avoid progression to cirrhosis and cancer and subsequently reduce morbidity and mortality.

**Acknowledgments**

The authors thank the NLST for the access to their data.

**Author Contributions**

Study design: Z.Z., J.W., R.Z., M.T.L., H.J.W.L.A.; code design, implementation, and execution: Z.Z., R.Z.; acquisition, analysis or interpretation of data: Z.Z., J.W., R.Z.; image annotation: J.W., J.T; writing of the manuscript: Z.Z, J.W., R.Z., J.T., H.J.W.L.A.; critical revision of the manuscript for important intellectual content: all authors; statistical analysis: Z.Z., J.W., R.Z.; study supervision: M.T.L., H.J.W.L.A.

**Competing Interests**

The authors declare no competing interests related to this work.

# SUPPLEMENTS

*Deep learning model architecture.* The deep learning model in the architecture of UNet (Supplemental Fig. 1) was developed for volumetric liver segmentation. The input dimension of the model was $64^3$, and the output dimension was $2 * 64^3$ since the output consists of background and target voxels. The MONAI platform integrated built-in packages which could map the model output to the CT image space. The output size of the segmentation function was equal to the input CT image size. Convolutional layers with the stride of 2 were utilized to increase/decrease each stage's dimensions. The kernel's size was 3*3*3 in all of the convolutional layers. In addition, parametric rectified linear units (PRelu) and batch normalization were introduced in each down-sampling and up-sampling step.

**Supplemental Figure 1:** The model architecture of the 3D U-Net for volumetric liver segmentation. **PRelu:** parametric rectified linear unit, **BN:** batch normalization.

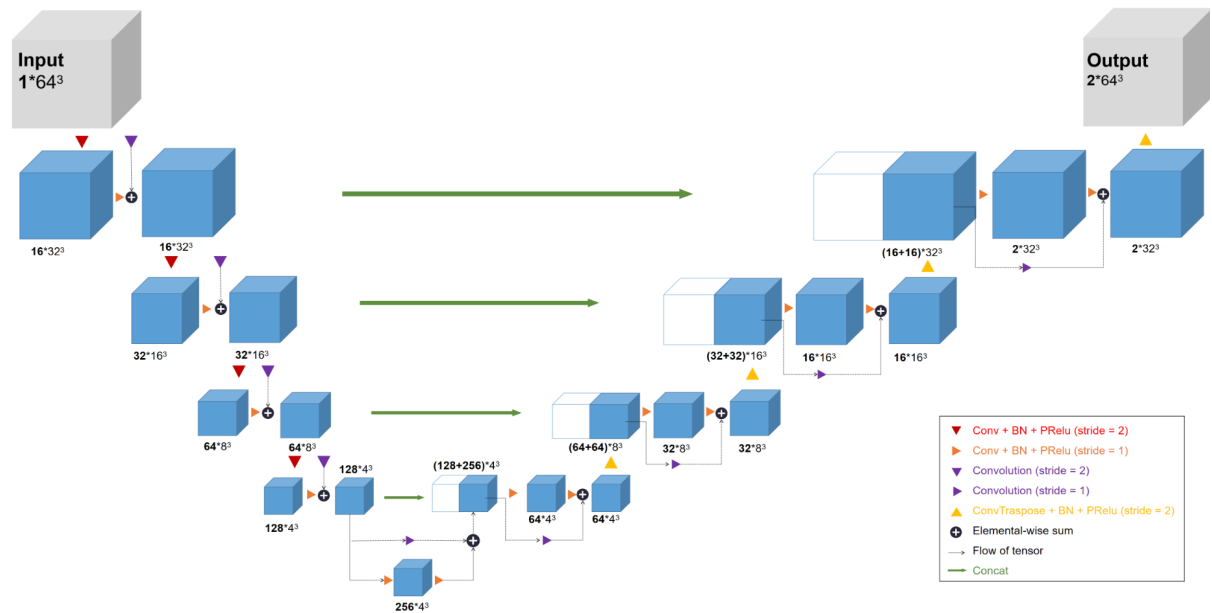

The main function of implementation is based on the MONAI 0.4.0 and Pytorch 1.6.0. No augmentation techniques were implemented thanks to the vast amount of training data. The parameters of the model were updated with the Adam optimizer at a learning rate of 1e-4. Starting parameter of PRelu is 0.25. The loss function is the Dice loss function. No transfer learning is adopted. The initial learning rate was 1e-4 with 0.1 decaying rates at every 150 epochs, during which the optimal model was taken as the final model. In total, 450 training epochs took 32 hours. The GPU is an Nvidia Titan RTX with 24GB memories.

*Incorrect segmentation of deep learning.* To analyze the outliers of the deep learning system for volumetric liver segmentation, we presented the worst examples of incorrectly segmented cases in the test set (n=100). In Supplemental Figure 2, CT example 1 has the lowest dice score and the biggest discrepancy of the liver volume; CT example 2 has the second-worst dice score and liver volume discrepancy; CT example 3 has the biggest discrepancy of the liver attenuation and the fourth-worst dice score.

**Supplemental Figure 2:** Incorrectly segmented cases of deep learning pipeline. In Figures A and B, Spearman correlation coefficients with associated two-sided p-value were estimated for the comparison of expert readers and artificial intelligence measurements in HU and liter. For Figure C, each image was sliced from the CT scan with the AI (blue) and expert reader (green) segmentation masks.

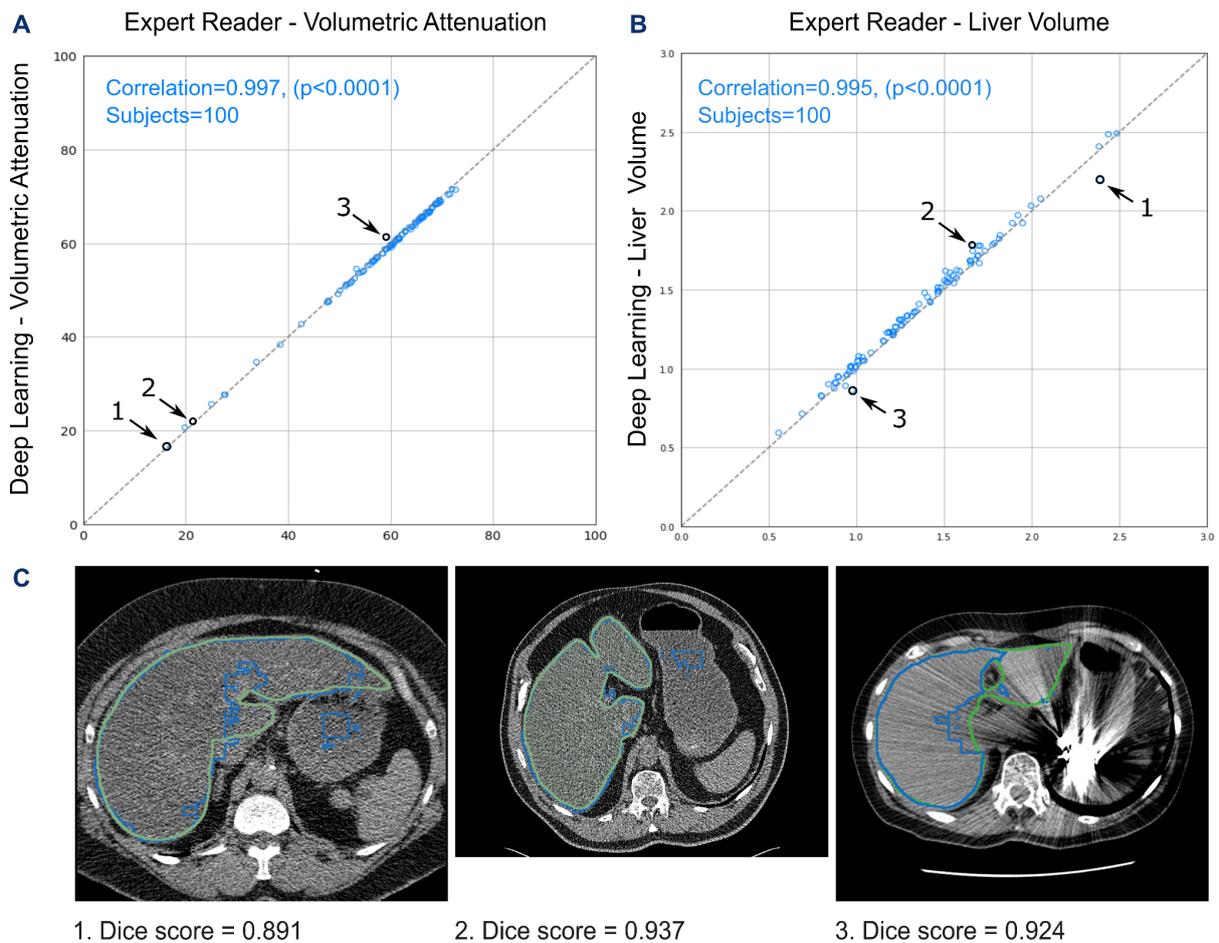